\documentclass[12pt]{article}

\usepackage[]{graphicx}
\usepackage{fullpage}
\usepackage{times}

\title{The Effects of Weak Spatiotemporal Noise on a\\
Bistable One-Dimensional System}

\author{Robert S. Maier$^{(a,b)}$ and D.~L. Stein$^{(b,a)}$\\
Depts.\ of $^{(a)}$Mathematics and $^{(b)}$Physics\\
University of Arizona\\
Tucson, AZ 85721, USA}

\date{}

\begin{document} 
\maketitle 

\begin{abstract}
We treat analytically a model that captures several features of the
phenomenon of spatially inhomogeneous reversal of an order parameter.  The
model is a classical Ginzburg--Landau field theory restricted to a bounded
one-dimensional spatial domain, perturbed by weak spatiotemporal noise
having a flat power spectrum in time and space.  Our analysis extends the
Kramers theory of noise-induced transitions to the case when the system
acted on by the noise has nonzero spatial extent, and the noise itself is
spatially dependent.  By extending the Langer--Coleman theory of the
noise-induced decay of a metastable state, we determine the dependence of
the activation barrier and the Kramers reversal rate prefactor on the size
of the spatial domain.  As this is increased from zero and passes through a
certain critical value, a transition between activation regimes occurs, at
which the rate prefactor diverges.  Beyond the transition, reversal
preferentially takes place in a spatially inhomogeneous rather than in a
homogeneous way.  Transitions of this sort were not discovered by Langer or
Coleman, since they treated only the infinite-volume limit.  Our analysis
uses higher transcendental functions to handle the case of finite volume.
Similar transitions between activation regimes should occur in other models
of metastable systems with nonzero spatial extent, perturbed by weak noise,
as the size of the spatial domain is varied.
\end{abstract}



\section{Introduction}
\label{sec:intro}

The phenomenon of noise-induced escape from a metastable state, and the
related phenomenon of noise-induced transitions between the stable states
of a bistable system, occur in many places in the sciences and
engineering~\cite{McClintock89a}.  Our focus in this paper is the case when
the system on which the noise acts has nontrivial spatial extent, and the
noise is not spatially uniform, but has significant spatial dependence.
This case arises in numerous physical contexts, in both equilibrium and
non-equilibrium systems.  Examples of noise-induced transitions in
spatially extended systems include pattern formation in convective and
electroconvective systems far from equilibrium~\cite{Cross93,Bisang98},
thermally activated magnetization reversal in
nanomagnets~\cite{Braun93,Brown00}, vortex creep in
superconductors~\cite{Blatter94}, the growth of instabilities in metallic
nanowires~\cite{Buerki2003}, and many others.

Possibly the simplest mathematical model of a spatially extended system
undergoing a noise-induced transition is a Ginzburg--Landau scalar field
theory perturbed by weak spatiotemporal noise.  Even here, the mathematical
complexity of the problem has restricted most treatments to the case of one
spatial dimension.  The phenomenon of classical nucleation on a line was
treated by Langer~\cite{Langer69} in a seminal paper, and a closely related
quantum phenomenon was treated by Callan and Coleman~\cite{Callan77}.  In~the
limit as the size of the one-dimensional spatial domain tends to infinity,
they were able to compute (respectively) the nucleation rate per unit
length, and what amounts to a tunnelling rate per unit length.  Their work
has been widely applied and cited.

However, many practical applications require an understanding of the
phenomenon of noise-induced nucleation in systems with a spatial extent
that is both nonzero and non-infinite, and the extent to which the
nucleation rate per unit length depends on system size.  Noise-induced
transitions between the two stable states of a bistable nonlinear field
theory, in a finite-volume domain, have been investigated by several
authors~\cite{Faris82,Martinelli89,McKane95}.  The difference between
finite and infinite systems is not merely quantitative.  Recently, the
authors~\cite{Maier02} uncovered an unusual effect, akin to a phase
transition, that occurs in an overdamped classical Ginzburg--Landau field
theory with a bistable $\phi^4$ potential, as the length~$L$ of its
one-dimensional spatial domain is varied.  A~similar phenomenon had
previously been seen in a quantum context~\cite{Chudnovsky92,Kuznetsov97}.
The lowest-energy `saddle' between the two stable configurations, through
which noise-activated transitions preferentially occur (especially when the
noise strength is low) may bifurcate.  Below a critical length~$L_c$, this
transition state is a spatially constant field configuration, but
at~$L=L_c$ it bifurcates into a spatially varying pair of configurations,
degenerate in energy.  Subsequent work~\cite{Stein2003} shows that a similar
bifurcation occurs in a Ginzburg--Landau model with an asymmetric
potential, in which the two new spatially varying transition states are not
degenerate; so one or the other is preferred.  The `phase transition' at a
critical length $L_c$ is therefore reasonably robust.

So at a critical system size, there may occur a major change in the
phenomenology of noise-induced transitions between the stable states of a
Ginzburg--Landau field theory, which proceed via nucleation.  Associated
with the bifurcation of the transition state is a bifurcation of the MPEP
(most probable escape path, i.e., transition path in configuration space)
extending uphill from each stable state to the transition state.  The
bifurcation is driven by this preferred nucleation pathway becoming
unstable in the transverse direction.  As~one would expect, the transition
rate is strongly affected by the bifurcation.  Formally, the prefactor in
the Kramers (weak-noise) nucleation rate {\em diverges\/} at~$L=L_c$.  This
signals that precisely at $L=L_c$, the phenomenon of transition between the
two stable states becomes non-Arrhenius: the rate at which it occurs falls
off in the limit of weak noise not like an exponential (with a constant
prefactor), but rather like an exponential with a power-law prefactor.

Our paper Ref.~\cite{Maier02} computed the Kramers prefactor as a function
of~$L$, through the bifurcation, by using elliptic functions.  The use of
higher transcendental functions made it possible to treat the case
$L<\infty$, i.e., to go beyond the analyses of Langer, and Callan and
Coleman.  However, we considered in~detail only the case when Dirichlet
conditions are imposed on the Ginzburg--Landau field at the endpoints of
the spatial domain.  How sensitive is the occurrence of a phase transition
at some $L=L_c$ to the boundary conditions employed?  We examine this
question here, by extending our qualitative results and nucleation rate
computations to the cases of Neumann, periodic, and antiperiodic boundary
conditions.  We~shall show that the phase transition is robust to changes
in boundary condition, but that unsurprisingly, there are quantitative
differences among the various cases.  Examination of these differences will
permit us to present our methods in more detail, and provide further
insights into the behavior of noise-activated transitions in spatially
extended systems.

\section{Model and Phenomenology}
\label{sec:model} 

\begin{figure}
\begin{center}
\begin{tabular}{c}
\includegraphics[height=5cm]{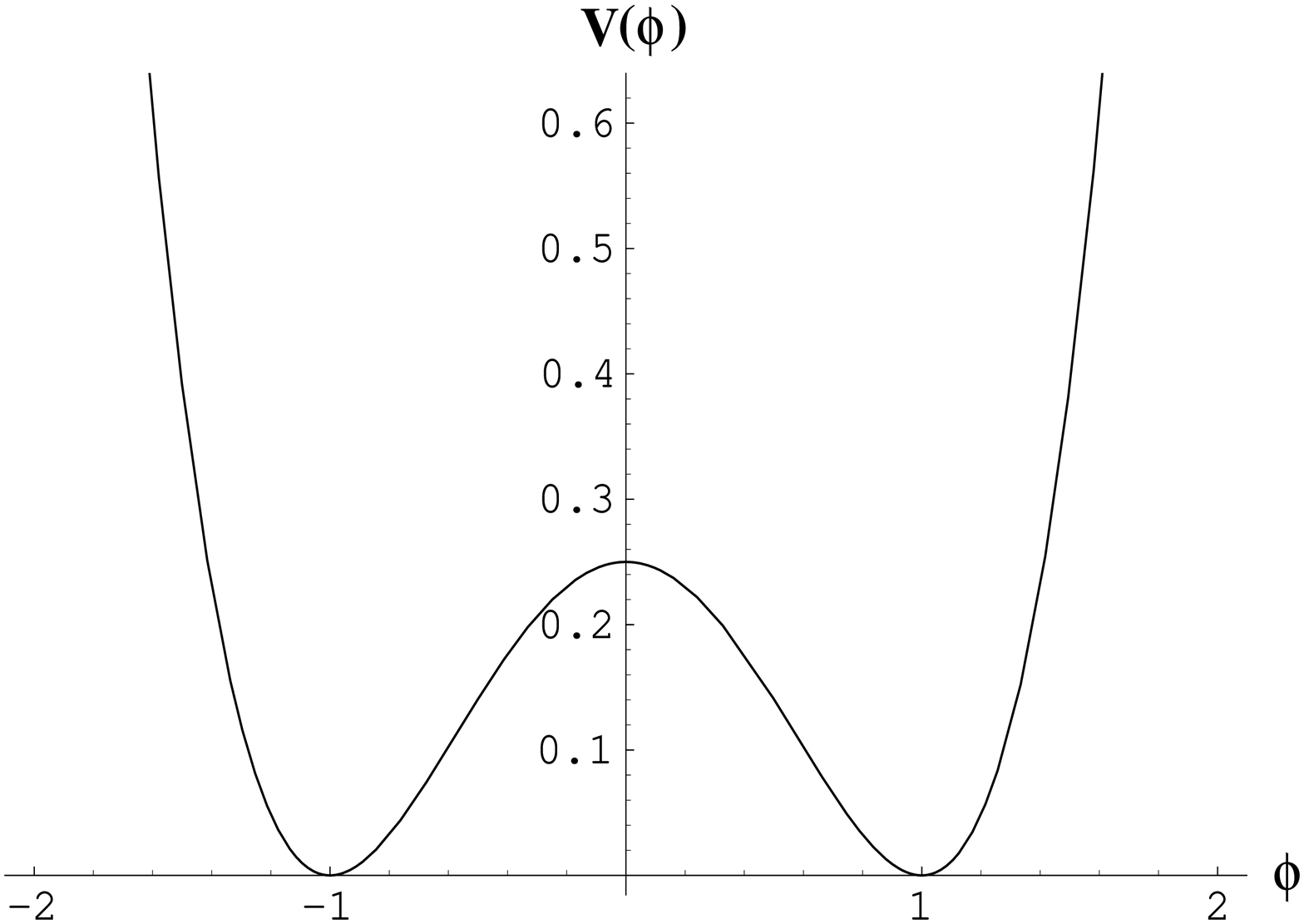}
\end{tabular}
\end{center}
\caption[The potential.]{\label{fig:potential} 
Bistable potential for the order parameter~$\phi$ given by Eq.~\ref{eq:field}.}
\end{figure} 

Consider an order parameter~$\phi$ governed by a bistable quartic potential
energy function that may be written in dimensionless units as
\begin{equation}
\label{eq:field}
V(\phi) = (\phi^2-1)^2/4 \, .
\end{equation}
This potential is shown in Fig.~\ref{fig:potential}.  The stable states
$\phi=\pm1$ are degenerate, with zero energy.  If $\phi$ evolves in a
deterministic, overdamped way, it will satisfy the equation
$\dot\phi=-V'(\phi)$.  If~it is perturbed by additive white noise, it will
satisfy, instead, $\dot\phi = -V'(\phi)+\epsilon^{1/2}\xi(t)$.  Here
$\xi$~is unit-strength temporal white noise, satisfying $\langle
\xi(t_1)\xi(t_2)\rangle=\delta(t_1-t_2)$, and $\epsilon$ is the noise
strength.  In a thermal context, $\epsilon\propto kT$.

We shall treat here the case when $\phi$~is actually a classical field on a
one-dimensional spatial domain: in dimensionless units, the interval
$[0,L]$.  The simplest extension of the overdamped stochastic evolution
equation to incorporate spatial effects is the stochastic Ginzburg--Landau
equation
\begin{eqnarray}
\label{eq:GL}
\dot\phi&=&\phi'' -V'(\phi) + \epsilon^{1/2}\xi(x,t)\\
&=&\phi'' + \phi - \phi^3 + \epsilon^{1/2}\xi(x,t)\,,
\nonumber
\end{eqnarray}
where $\xi(x,t)$ is unit-strength spatiotemporal white noise, which is
defined to satisfy
$\langle\xi(x_1,t_1)\xi(x_2,t_2)\rangle=\delta(x_1-x_2)\delta(t_1-t_2)$.
Zero-noise dynamics will be `gradient', i.e., conservative.  That~is, if
$\epsilon=0$, then
\begin{equation}
\label{eq:gradient}
\dot\phi=-\delta{\cal H}/\delta\phi\, ,
\end{equation}
where
\begin{equation}
\label{eq:energy}
{\cal H}[\phi]\equiv
\int_{0}^{L} \left[\frac12(\phi')^2 +V(\phi)\right]
\,dx
\end{equation}
is the energy functional.  So the statistical properties of the
stochastically evolving field~$\phi$ are described by equilibrium
statistical mechanics.  However, just as in the model without spatial
extent, nonzero noise can induce transitions between the stable states,
which naively are the two field configurations $\phi\equiv\pm1$.  This will
typically occur via nucleation.  A~droplet of one stable configuration will
form in a background of the other, and will `nucleate': under the driving
influence of the noise, it will spread to fill the entire spatial domain.
Of~course, it is more likely for a small droplet to shrink and vanish,
especially when the noise is weak.

In the infinite-dimensional configuration space, an MPEP goes `uphill' from
each stable field configuration, leading to a preferred transition
configuration (saddle) that lies between them.  Each of these MPEPs is a
nucleation pathway, i.e., a path of least resistance.  By time-reversal
invariance, each MPEP is a time-reversed zero-noise `downhill' trajectory.
If the noise is weak, the order parameter is expected to flip between the
two stable configurations in a Markov way, with the expected waiting time
in the basin of attraction of each being an exponential random variable, as
is typical of slow rate processes.  The activation rate (the reciprocal of
the mean time between flips) will be given in the $\epsilon\to0$ limit by
the Kramers formula
\begin{equation}
\label{eq:Kramers}
\Gamma\sim \Gamma_0\exp(-\Delta W/\epsilon)\, .
\end{equation}
Here $\Delta W$ is the activation barrier, which quantifies the extent to
which the preferred transition configuration between the two stable
configurations is energetically disfavored, and $\Gamma_0$ is the rate
prefactor.  Due to the normalization convention implicit in~(\ref{eq:GL}),
$\Delta W = 2\Delta E$, where $\Delta E$ is the energy of the transition
state minus the energy of either stable state.  Here energy is computed
from the functional~$\cal{H}[\cdot]$ of~(\ref{eq:energy}).  The factor
of~$2$ arises from our decision to multiply unit-strength spatiotemporal
noise by~$\epsilon^{1/2}$ rather than by~$\sqrt{2\epsilon}$.

The calculation of $\Delta W$ and $\Gamma_0$ will make up the technical
component of this paper.  Since we are using dimensionless units, they will
depend only on the length~$L$ and the choice of boundary conditions at the
endpoints $x=0$ and~$x=L$.  The boundary conditions affect the way in which
order parameter reversal occurs, since they may force nucleation to begin,
preferentially, at the endpoints.  It~should also be noted that for some
choices of boundary conditions, the stable states may only be
approximations to the uniform $\phi\equiv\pm1$ configurations.

\section{The Stable and Transition States}
\label{sec:states}

We begin by examining the stationary solutions of the noiseless
($\epsilon=0$) evolution equation.  These will depend on the interval
length $L$ and the boundary conditions.  The four boundary conditions we
consider are periodic (P), antiperiodic (AP), Dirichlet (D), and Neumann
(N)\null.  All four, applied to the stochastic Ginzburg--Landau equation,
have potential applicability in physical modelling.  For example, in
modelling thermally activated magnetization reversal in a finite-length
ferromagnetic nanowire, Neumann boundary conditions are
appropriate~\cite{Braun99}.  The original treatments of Langer and
Callan--Coleman used periodic boundary conditions; and so forth.

\medskip

\noindent {\em Periodic.}---The conditions are $\phi(0)=\phi(L)$,
$\phi'(0)=\phi'(L)$.  There are three constant time-independent solutions:
$\phi\equiv\pm 1$, each with energy 0; and $\phi\equiv0$, with
energy~$1/4$.  It is easy to see that $\phi\equiv\pm 1$ are stable for
any~$L$, and $\phi\equiv0$ is always unstable.  However, nonconstant
solutions may also exist for a range of~$L$.  These will be considered
below.

\noindent {\em Antiperiodic.}---The conditions are $\phi(0)=-\phi(L)$,
$\phi'(0)=-\phi'(L)$.  The only constant time-independent solution is
$\phi\equiv0$.

\noindent {\em Dirichlet.}---Only the conditions $\phi(0)=\phi(L)=0$ will
be considered here.  Many other Dirichlet boundary conditions could be
examined, but this case is particularly interesting because when $L$~is
sufficiently large, it~has two nonconstant stationary solutions, which
permits the possibility of a noise-induced transition.  This is the case
that was treated in Ref.~\cite{Maier02}.

\noindent {\em Neumann.}---Though there are many possibilities, we consider
here only the conditions $\phi'(0)=\phi'(L)=0$.  The constant solutions
here are the same as for periodic boundary conditions, but it will be seen
that there are important differences between the two cases.

What {\em nonconstant\/} time-independent solutions exist when
$\epsilon=0$?  By~(\ref{eq:GL}), any stationary solution satisfies
\begin{equation}
\label{eq:EL}
\phi''=-\phi+\phi^3\,,
\end{equation}
subject to the given boundary condition.  That is, $\dot\phi=-\delta{\cal
H}/\delta\phi=0$.  The linearized noiseless dynamics in the vicinity of
such a state are specified by the Hessian operator $\delta^2{\cal
H}/\delta\phi^2$.  Stationary configurations for which this operator has
all positive eigenvalues are stable; those with a single negative
eigenvalue are possible transition states.  For the latter, the eigenvector
corresponding to the negative eigenvalue is the direction along which the
MPEP approaches, in the infinite-dimensional configuration space.
Nonconstant transition states are often called {\em instanton states\/}, in
a nomenclature derived from Callan and Coleman~\cite{Callan77}.  In
$\phi^4$ and other theories, the term `instanton' usually refers to a
kink-like field configuration $\phi=\phi(x)$ asymptotic to $\phi=\pm1$ as
$x\to\pm\infty$, with a single node~\cite{Rajaraman82}.

To satisfy the specified boundary conditions on~$[0,L]$, a rather different
sort of instanton state must be used.  It~is easy to check that the
so-called {\em periodic instanton\/} $\phi=\phi_{{\rm inst},m}(x)$ is a
time-independent solution of~(\ref{eq:GL}) for any~$m$ in the range
$0<m\le1$.  Here (cf.\ Ref.~\cite{Espichan00}),
\begin{equation}
\label{eq:instanton}
\phi_{{\rm inst},m}(x) \equiv
\sqrt{\frac{2m}{m+1}}\,{\rm sn}(x/\sqrt{m+1}\mid m)\,,
\end{equation}
where ${\rm sn}(\cdot\mid m)$ is the Jacobi elliptic $\rm sn$ function with
parameter~$m$.  Its quarter-period is given by~${\bf K}(m)$, the complete
elliptic integral of the first kind~\cite{Abramowitz65}, which is a
monotonically increasing function of~$m$.  It can be viewed as an infinite
alternating sequence of kinks and anti-kinks, spaced a distance $2{\bf
K}(m)$ apart.  As~$m\to 0^+$, ${\bf K}(m)$ decreases to~$\pi/2$, and ${\rm
sn}(\cdot\mid m)$ degenerates to $\sin(\cdot)$.  As $m\to 1^-$, the
quarter-period increases to infinity (with a logarithmic divergence), and
${\rm sn}(\cdot\mid m)$ degenerates to the nonperiodic function
$\tanh(\cdot)$, which is the canonical single-kink sigmoidal function.
It~is no accident that the hyperbolic tangent function appeared in the
Langer and Callan--Coleman analyses in connection with the limiting
($L\to\infty$) shape of a critical droplet.  For a careful recapitulation,
see Ref.~\cite{Schulman}.

In the present context, the value of~$m$ in~(\ref{eq:instanton}) is
determined by the interval length $L$ and the boundary conditions, as
follows.

\medskip

\noindent {\em Periodic.}---As stated earlier, the uniform configurations
$\phi\equiv\pm 1$ are the stable states for all~$L$.  If the nonuniform
solution~(\ref{eq:instanton}) is to satisfy the boundary conditions, then
$L$~must be an integer multiple of a full period.  It~is obvious that
larger integers correspond to higher energies (and therefore activation
barriers); so the physically relevant nonuniform transition state is
$\phi=\phi_{{\rm inst},{m_{u,P}}}(x)$, where $m_{u,P}$ is determined
implicitly by
\begin{equation}
\label{perm}
4\sqrt{m_{u,P}+1}\ {\bf K}(m_{u,P})=L\,.
\end{equation}
The subscript~$u$ denotes an unstable (i.e., transition) state and
$P$~denotes periodic boundary conditions.  Similarly, $s$~will denote a
stable state, and the other boundary conditions will be correspondingly
abbreviated.

This nonuniform stationary state is not present for all~$L$: there is a
minimum value of~$L$, corresponding to $m=0$, below which no~solution of
the form $\phi=\phi_{{\rm inst},{m_{u,P}}}(x)$ can be constructed.  That
is, because ${\bf K}(0)=\pi/2$, the nonuniform transition state cannot
occur below $L=2\pi$.  Below this value, the only possible transition state
is the uniform state $\phi\equiv0$.  This solution remains an unstable
stationary state for $L>2\pi$, but with higher energy than that of the
instanton state.

These considerations suggest the following qualitative picture of the
periodic-b.c.\ case, to be justified subsequently by an energy and
eigenvalue analysis.  Suppose the system is in one of the two stable states
$\phi\equiv\pm1$.  How does a transition induced by (weak) noise to the
other state proceed?  If $L<2\pi$, the system preferentially `climbs
uphill' to the $\phi\equiv0$ state, after which it `rolls downhill' to the
other stable state.  So reversal of the order parameter preferentially
takes place in a spatially homogeneous way.  If $L>2\pi$, however, there is
another saddle, lower in energy than the instanton state and therefore a
much more probable intermediate state when the noise strength is small.
This is the instanton configuration $\phi=\phi_{{\rm inst},{m_{u,P}}}(x)$,
which is positive on half the interval and negative on the other half.
This configuration can be thought of as a `droplet pair', in which half the
spatial domain is occupied, to a first approximation, by each of the two
stable values of~$\phi$.  Equivalently, it can be viewed as a configuration
with a kink and an anti-kink, or two Bloch walls (if~one is modelling a
magnetic system).  A~transition occurs in the following way: a~small
droplet of the reversed value for the order parameter is formed, and under
the driving influence of the noise, it~grows to occupy occupy half the
interval.  At~that point, the saddle has been reached, and the droplet
continues growing of its own accord until it fills the remainder of the
interval, i.e., `rolls downhill'.  Both halves of the transition can be
interpreted in~terms of kink motion: the uphill half being noise-driven,
and the downhill half being deterministic.

Actually, there is a complication here: due to the translation symmetry
that accompanies periodic boundary conditions, the transition state is
necessarily infinitely degenerate.  That~is, $\phi=\phi_{{\rm
inst},{m_{u,P}}}(x-x_0)$, for any~$x_0$, will serve as a transition state.
In physical language, the kink--anti-kink pair, i.e., the droplet, may form
anywhere along the interval, leading (in~the $L\to\infty$ limit) to a
transition rate {\em per unit length\/}.  We~treat this matter
elsewhere~\cite{Maier08} (see also Ref.~\cite{Stein2003}).

\noindent {\em Antiperiodic.}---In this case there is a nonuniform stable
state of the form $\phi=\phi_{{\rm inst},m_{s,AP}}$, with $m_{s,AP}$
defined implicitly by
\begin{equation}
\label{eq:antiperms}
2\sqrt{m_{s,AP}+1}\ {\bf K}(m_{s,AP})=L\, , 
\end{equation}
However, this equation has no solution when $L<\pi$; in this range there is
a single stable state given by $\phi\equiv0$, and no possibility of an
accompanying transition.  The $\phi\equiv0$ solution becomes unstable when
$L>\pi$, and in fact is the transition state for $\pi<L<3\pi$.  For
larger~$L$ this transition state undergoes a pitchfork bifurcation into a
pair of instanton states with {\em three\/} half-wavelengths in the
interval $[0,L]$.  That~is,
\begin{equation}
\label{eq:antipermu}
6\sqrt{m_{u,AP}+1}\ {\bf K}(m_{u,AP})=L\, .
\end{equation}
As in the periodic case, this nonuniform transition state is infinitely
degenerate; and the nonuniform stable state is~too.

\noindent {\em Dirichlet.}---This case was examined in detail in
Ref.~\cite{Maier02}.  If $L\le\pi$, the model is monostable with the
configuration $\phi\equiv0$ as the only stable state.  When $L$~increases
through~$\pi$, this state undergoes a pitchfork bifurcation into a pair of
stable configurations with
\begin{equation}
\label{dirs}
2\sqrt{m_{s,D}+1}\ {\bf K}(m_{s,D})=L\, .
\end{equation}
In the range $0\le L\le 2\pi$, the $\phi\equiv0$ configuration is the
transition state.  It undergoes a pitchfork bifurcation at $L=2\pi$; for
larger values of~$L$ the parameter~$m$ of the transition state is
determined by
\begin{equation}
\label{eq:diru}
4\sqrt{m_{u,D}+1}\ {\bf K}(m_{u,D})=L\, .
\end{equation}
Note that the Dirichlet condition (and the Neumann condition, to be
discussed below), pin the nonuniform stationary states at the boundaries,
so that there is no issue of infinite degeneracy arising from translation
symmetry to complicate the analysis.  In physical terms, each of the two
preferred nucleation pathways begins with the formation of a droplet at
either end of the interval.  As $L\to\infty$, it is not natural to speak of
a transition rate per unit length.

\begin{figure}
\begin{center}
\begin{tabular}{c}
\includegraphics[height=6.25cm]{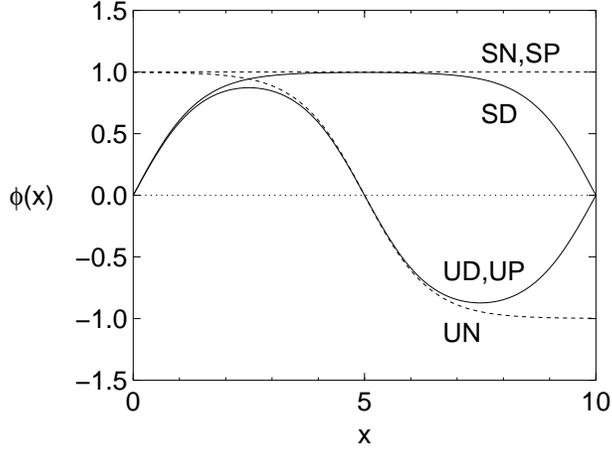}
\end{tabular}
\end{center}
\caption[The stationary states.]{\label{fig:sols} The stable states~(S) and
transition states~(U), for periodic~(P), Dirichlet~(D), and Neumann~(N)
boundary conditions, when $L=10$.  Each nonzero state has a degenerate
counterpart obtained by $\phi\mapsto -\phi$, and UP~may be shifted
arbitrarily, so it is infinitely degenerate.  (From
Ref.~\cite{Maier02}.)}
\end{figure}

\noindent {\em Neumann.}---Here the stable states are the uniform states
$\phi\equiv\pm 1$.  Also, if $L\le\pi$ the transition state is the uniform
configuration $\phi\equiv0$.  When $L$~is increased through~$\pi$, the
transition state bifurcates into a pair of instanton states with the first
argument of the Jacobi elliptic $\rm sn$ function in~(\ref{eq:instanton})
shifted by ${\bf K}(m)$, a quarter wavelength (equivalent here to $L/2$).
That is, the transition state is given by
\begin{equation}
\label{eq:neuinst}
\sqrt{\frac{2m}{m+1}}\,{\rm sn}(x/\sqrt{m+1}+{\bf K}(m)\mid m)\, ,
\end{equation}
with $m=m_{u,N}$, a quantity determined implicitly by
\begin{equation}
\label{eq:neuu}
2\sqrt{m_{u,N}+1}\ {\bf K}(m_{u,N})=L\, .
\end{equation}
So $m_{u,N}=m_{s,D}$.  In fact, up~to a uniform shift, the Neumann-case
transition state is the same as the Dirichlet-case stable state.

\smallskip

The results so far are summarized in Fig.~\ref{fig:sols}, which depicts the
stable and transition states for the cases of periodic, Dirichlet, and
Neumann boundary conditions (the antiperiodic case is excluded for figure
clarity).  Each plotted nonconstant curve consists of an integer number of
kinks or anti-kinks; or more accurately, an even number of half-kinks or
half-anti-kinks.

\section{The Activation Barrier}
\label{sec:action}

As mentioned in Sec.~\ref{sec:model}, the exponential falloff of the transition
rate in the limit of weak noise, i.e., its
Arrhenius behavior, is determined by the activation barrier $\Delta W$
between the two stable states.  We~noted that $\Delta W=2\Delta E=2({\cal
H}[\phi_u]-{\cal H}[\phi_s])$, with the energy functional ${\cal H}[\phi]$
given by~(\ref{eq:energy}).  We~shall call $\Delta E$ the `activation
energy'.  The calculation of ${\cal H}[\phi_s]$ and ${\cal H}[\phi_u]$, the
stable and transition state energies, is trivial in the case of uniform
states, and reasonably straightforward in the case of nonuniform
(instanton) states, if standard elliptic function
formulas~\cite{Abramowitz65} are used.  The results are as follows.

\medskip

\noindent {\em Periodic.\/}---If $L<L_c^P=2\pi$, then trivially, $\Delta
E=L/4$.  If $L>L_c^P$, then
\begin{equation}
\label{eq:actp}
\Delta E = {1\over 3\sqrt{1+m_{u,P}}}\left[8{\bf
E}(m_{u,P})-{(1-m_{u,P})(3m_{u,P}+5)\over (1+m_{u,P})}{\bf
K}(m_{u,P})\right]\, ,
\end{equation}
where ${\bf E}(m)$ is the complete elliptic integral of the second
kind~\cite{Abramowitz65}.  Note that as $m_{u,P}\to 0^+$, i.e.,
$L\to(2\pi)^+$, the two regions connect in a smooth manner: $\Delta E$~is
is differentiable with respect to~$L$ at $L=2\pi$, though not
twice differentiable (the same relatively smooth join will obtain in all
succeeding cases).

As $m_{u,P}\to 1^-$ (i.e., $L\to\infty$), $\Delta E\to 4\sqrt{2}/3$.  This
is a familiar quantity: it~is twice the energy of a $\phi^4$
kink~\cite{Rajaraman82}.  The factor of~$2$ is due to the transition state
configuration~$\phi_u$, the stationary periodic instanton solution, having
two nodes, i.e., two swings between $\phi=\pm1$ as $x$~varies from $0$
to~$L$.  In field theory language, it comprises a kink and an anti-kink.

\smallskip

\noindent {\em Antiperiodic.\/}---In the range $\pi<L\le 3\pi$,
\begin{equation}
\label{eq:actap-}
\Delta E = {2\over 3(1+m_{s,AP})^{3/2}}\Big[(m_{s,AP}+2){\bf K}(m_{s,AP})-2(m_{s,AP}+1){\bf
E}(m_{s,AP})\Big]\, .
\end{equation}
If $L>L_c^{AP}=3\pi$, then
\begin{eqnarray}
\label{eq:actap+}
\Delta E &=& {4\over 3}\left[{3\over\sqrt{1+(m_{u,AP})}}{\bf E}(m_{u,AP}) - {1\over
\sqrt{1+(m_{s,AP})}}{\bf E}(m_{s,AP})\right]\\
&&\quad{}-{1\over 6}\left[{3(1-m_{u,AP})\over(1+m_{u,AP})^{3/2}}(3m_{u,AP}+5){\bf K}(m_{u,AP})
-{1-m_{s,AP}\over(1+m_{s,AP})^{3/2}}(3m_{s,AP}+5){\bf K}(m_{s,AP})\right]\nonumber\,.
\end{eqnarray}
It is noteworthy that as $m_{u,AP}\to 1^-$, $\Delta E\to 4\sqrt{2}/3$, just
as in the periodic case.  In the periodic case, the transition state has
two nodes while the stable state has zero nodes; in the antiperiodic case,
the transition state has three nodes while the stable state has one node.
It~is therefore not surprising that in both cases, the energy difference
converges to twice the energy of a $\phi^4$ kink as $L\to\infty$.  Note
that for any finite~$L$, the relevant $m$ values differ between the
periodic and antiperiodic cases, but all approach unity from below
as~$L\to\infty$.

\smallskip 

\begin{figure}
\begin{center}
\begin{tabular}{c}
\includegraphics[height=6.25cm]{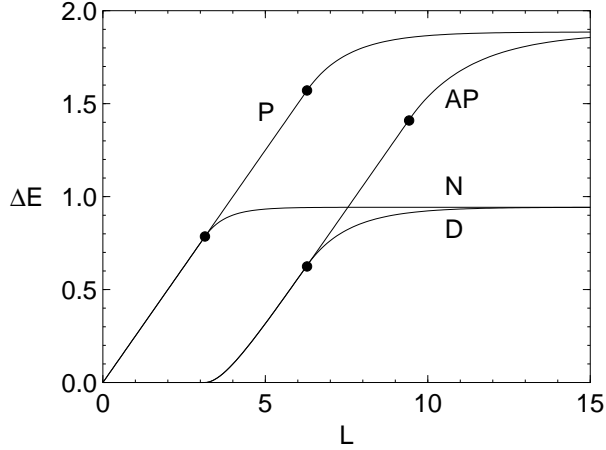}
\end{tabular}
\end{center}
\caption[The activation energy.]{\label{fig:acten} The activation energy
$\Delta E$ as a function of the interval length~$L$, for periodic (P),
antiperiodic (AP), Dirichlet (D), and Neumann~(N) boundary conditions.
Bullets indicate critical interval lengths, at which bifurcations take
place ($L^N_c=\pi$; $L^D_c=L^P_c=2\pi$; and $L^{AP}_c=3\pi$).}
\end{figure}

\noindent {\em Dirichlet.\/}---If $L\le L_c^D=2\pi$, then
\begin{equation}
\label{eq:actd-}
\Delta E = {2\over 3(1+m_{s,D})^{3/2}}\Big[(m_{s,D}+2){\bf K}(m_{s,D})-2(m_{s,D}+1){\bf
E}(m_{s,D})\Big]\, ,
\end{equation}
which has the same form as~(\ref{eq:actap-}).  If $L>L_c^D$, then
\begin{eqnarray}
\label{eq:actd+}
\Delta E &=& {4\over 3}\left[{2\over\sqrt{1+(m_{u,D})}}{\bf E}(m_{u,D}) -
{1\over\sqrt{1+(m_{s,D})}}{\bf E}(m_{s,D})\right]\\
&&\quad{}-{1\over 6}\left[{2(1-m_{u,D})\over(1+m_{u,D})^{3/2}}(3m_{u,D}+5){\bf K}(m_{u,D})
-{1-m_{s,D}\over(1+m_{s,D})^{3/2}}(3m_{s,D}+5){\bf
K}(m_{s,D})\right]\nonumber\,.
\end{eqnarray}
In the $L\to\infty$ limit, $\Delta E\to 2\sqrt{2}/3$, which is half the
value for the periodic and antiperiodic cases.  This is because the
transition state now has only one more node than the stable state, rather
than two.  The same is true of the Neumann case.  Hence these two latter
cases will have the same activation energy in the $L\to\infty$ limit, just
as the periodic and antiperiodic values for~$\Delta E$ converge in this
limit; and it will be the energy of a single $\phi^4$ kink.

\smallskip

\noindent {\em Neumann.\/}---If $L\le L_c^N=\pi$, then $\Delta E = L/4$;
both the stable and transition states are the same here as in the low-$L$
regime of the periodic case.  If $L>L_c^N$, then
\begin{equation}
\label{eq:actn}
\Delta E = {1\over 3(1+m_{u,N})^{3/2}}\left[4(1+m_{u,N}){\bf E}(m_{u,N})
-{1\over 2}(1-m_{u,N})(3m_{u,N}+5){\bf K}(m_{u,P})\right]\, ,
\end{equation}
As noted, the activation energy in the $L\to\infty$ limit equals
$2\sqrt2/3$, as in the Dirichlet case.

\medskip

The above formulas for the activation energy $\Delta E$ as a function of
$L$ are plotted in Fig.~\ref{fig:acten}.  The bifurcations at $L=\pi$,
$L=2\pi$, and $L=3\pi$ are apparent, as is the differentiability (and lack
of twice differentiability) through each bifurcation.

\section{The Transition Rate Prefactor}
\label{sec:prefactor}

As explained, the activation energy is (up~to a factor of~$2$) equal to the
activation barrier $\Delta W$ in the Kramers transition rate formula
$\Gamma\sim\Gamma_0\exp(-\Delta W/\epsilon)$, $\epsilon\to0$.  Calculation
of the prefactor $\Gamma_0$ is a much more involved matter.  In~general, it
requires an analysis of the {\em transverse fluctuations\/} about the MPEPs
that go uphill to the transition state.  In any `zero-dimensional' (i.e.,
non-spatially-extended) equilibrium system, in which the noise-perturbed
evolution equation is a stochastic {\em ordinary\/} differential equation,
it is known that it suffices to compute the eigenvalues of the linearized
dynamics at the endpoints of each MPEP, i.e., in the vicinity of the stable
states and transition state~\cite{Gardiner85}.  For example, a system with
a two-dimensional order parameter~${\phi}$ satisfying the overdamped
evolution equation $\dot{\phi}={\bf u}({\phi})+\epsilon^{1/2}{\bf\xi}$,
where ${\bf u}={\bf u}({\phi})$ is a drift field that is the negative
gradient of a potential, will have the Kramers escape rate
\begin{equation}
\label{eq:Eyring}
\Gamma\sim
{1\over2\pi}\sqrt{\left|\lambda_{\parallel}(U)\right|\lambda_{\parallel}(S)}\
\sqrt{\lambda_{\perp}(S)\over
\lambda_{\perp}(U)}\exp(-\Delta W/\epsilon)\,,\qquad\epsilon\to0\,.
\end{equation}
Here $S$ denotes the stable fixed point of ${\bf u}$ and $U$ the saddle
point over which escape from the basin of attraction of~$S$ preferentially
occurs.  $\lambda_\parallel(S)=-{\partial u_{S,1}/\partial \phi_{S,1}}$ is
the eigenvalue of the linearized negative drift field at~$S$ whose
corresponding eigenvector points along $\hat\phi_{S,1}$, the direction
locally parallel to the MPEP\null.  Similarly, $\lambda_\perp(U)=-{\partial
u_{U,2}/\partial \phi_{U,2}}$ is the eigenvalue of the linearized negative
drift field at~$U$ whose corresponding eigenvector points
along~$\hat\phi_{U,2}$, the direction locally perpendicular to the MPEP,
and so forth.  $\lambda_\parallel(U)$~is the eigenvalue of the linearized
negative drift field at~$U$ that corresponds to the unstable, or `downhill'
direction; it is negative.  Eq.~(\ref{eq:Eyring}), which is called the
Eyring formula, is a version of the Kramers rate formula that incorporates
transverse fluctuations.

In a spatially extended equilibrium system, where the evolution equation is
a stochastic {\em partial\/} differential equation (cf.~(\ref{eq:GL})),
Eq.~(\ref{eq:Eyring}) generalizes in a formally (though not
computationally) straightforward way~\cite{Langer69}.  Because of the
field-theoretic nature of the model, the linearized dynamics at each fixed
point have a countable infinity of eigenvalues.  The generalization of the
prefactor~$\Gamma_0$ of~(\ref{eq:Eyring}) is
\begin{equation}
\label{eq:prodeig}
\Gamma_0=\frac{\left|\lambda_0(\phi_u)\right|}{2\pi}
\sqrt{
\frac
{\prod_{n=0}^\infty
\lambda_n(\phi_s)}
{\prod_{n=0}^\infty\left|\lambda_n(\phi_u)\right|}
}
\,,
\end{equation}
where $\phi_s$ is the stable configuration and $\phi_u$~is the dominant
saddle.  Here $\lambda_0(\phi_u)$, which is the only negative eigenvalue
in the sets $\{\lambda_n(\phi_u)\}_{n=0}^\infty$ and
$\{\lambda_n(\phi_s)\}_{n=0}^\infty$, 
is the negative eigenvalue of the
linearization of the negative drift field $\delta {\cal H}/\delta\phi$ at
the transition state~$\phi_u$.  It~corresponds to the unstable direction.
It~is easy to see that in the case of a two-dimensional order parameter, in
which there are only two eigenvalues at each stationary state,
(\ref{eq:prodeig})~reduces to the Eyring formula~(\ref{eq:Eyring}).
$\lambda_\parallel(U)$ corresponds to~$\lambda_0(\phi_u)$.

Because both the numerator and denominator of the quotient inside the
square root are products of an infinite number of eigenvalues (typically
with magnitude much greater than unity, as will shortly be seen), they may
diverge.  However, their {\em ratio\/}, defined in a limiting sense, will
generally be finite.

It is often not possible to compute in closed form the eigenvalue spectrum
of the linearized zero-noise dynamics at the relevant stationary points
(although an exception will be seen below).  To~employ standard techniques
for prefactor computation, the above formula may optionally be recast as a
{\em determinant quotient\/} of the linearized time-evolution operators at
the fixed points~\cite{McClintock89a,Langer69,Schulman,Hanggi90}.
Consider a small perturbation ${\eta}$ about the stable state, i.e.,
${\phi}={\phi}_s+{\eta}$.  Then to leading order
$\dot{\eta}=-{\hat\Lambda}[\phi_s]{\eta}$, where ${\hat\Lambda}[\phi_s]$,
which is the Hessian operator $\delta^2 {\cal H}/\delta\phi^2$ evaluated at
$\phi=\phi_s$, specifies the linearized zero-noise dynamics at~${\phi}_s$.
Similarly, ${\hat\Lambda}[\phi_u]$ specifies the linearized zero-noise
dynamics at~${\phi}_u$.  For notational convenience, we shall write
$\lambda^s_n$ and~$\lambda^u_n$ for $\lambda_n(\phi_s)$
and~$\lambda_n(\phi_u)$.  As~above, let $\lambda_{0}^u$ be the only
negative eigenvalue of~${\hat\Lambda}[\phi_u]$, corresponding to the
direction along which the MPEP approaches the transition state.  Then
\begin{equation}
\label{eq:detrat}
\Gamma_0 = \frac{\left|\lambda_{0}^u\right|}{2\pi}
\sqrt{\frac{\det{\hat\Lambda}[\phi_s]}{\bigl|\det{\hat\Lambda}[\phi_u]\bigr|}}
\end{equation}
is an alternative expression for~$\Gamma_0$.  The two determinants are
typically not defined, but their quotient can be made sense~of in any of
several ways: as~a limit of determinant quotients arising from
finite-dimensional truncations, for instance.

In the stochastic Ginzburg--Landau model of~(\ref{eq:GL}), it is
straightforward to compute the differential operators $\hat\Lambda[\phi_s]$
and~$\hat\Lambda[\phi_u]$.  Linearizing the zero-noise evolution
$\dot\phi=-\delta {\cal H}/\delta\phi$ at any stationary state
$\phi=\phi_0$ yields
\begin{equation}
\label{eq:phi0}
\dot\eta = -\hat\Lambda[\phi_0]\,\eta \equiv
-\left[-{d^2}/{dx^2} + (-1 + 3\phi_0^2)\right]\eta\, .
\end{equation}
With the formulas (\ref{eq:detrat}) and~(\ref{eq:phi0}) in hand, we shall
now work~out the Kramers rate prefactor in the Neumann case, as a
illustrative calculation.

\subsection{The Neumann-Case Rate Prefactor When $L<L_c$}
\label{subsec:subcrit}

If $L<L_c^N=\pi$, both the stable and transition states are spatially
uniform: $\phi_s\equiv\pm1$ and $\phi_u\equiv0$.  This greatly simplifies
the computation of the associated eigenvalues.  Note that the eigenvalue
spectrum is the same at both stable states, by symmetry under
$\phi\mapsto-\phi$; and because only the square $\phi_0^2$ appears on the
right-hand side of~(\ref{eq:phi0}).

Linearizing around either stable state yields the operator
\begin{equation}
\label{eq:phis}
\hat\Lambda[\phi_s]=-d^2/dx^2+2\, ,
\end{equation}
and similarly
\begin{equation}
\label{eq:phiu}
\hat\Lambda[\phi_u]=-d^2/dx^2-1\, .
\end{equation}
The eigenvalue spectrum of $\hat\Lambda[\phi_s]$, equipped with Neumann
boundary conditions, is
\begin{equation}
\label{eq:stablespectrum}
\lambda_n^s=2+{\pi^2n^2\over L^2},\qquad\qquad\qquad n=0,1,2\ldots\,.
\end{equation}
The eigenvalue spectrum of $\hat\Lambda[\phi_u]$ is similarly
\begin{equation}
\label{eq:unstablespectrum}
\lambda_n^u=-1+{\pi^2n^2\over L^2},\qquad\qquad\qquad n=0,1,2\ldots\,.
\end{equation}
As expected, all eigenvalues of $\hat\Lambda[\phi_s]$ are positive, while
the $\hat\Lambda[\phi_u]$ operator has a single negative eigenvalue (with
value $-1$).  The corresponding eigenfunction, which is spatially uniform,
is the direction in configuration space along which the MPEP
approaches~$\phi_u$.

Putting everything together, we find the Neumann-case rate prefactor when
$L<L_c^N=\pi$ to be
\begin{eqnarray}
\label{eq:g0-}
\Gamma_0&=&{1\over2\pi}\,\sqrt{{\prod_{n=0}^\infty\bigl(2+{\pi^2n^2\over
L^2}\bigr)\over
\left|\prod_{n=0}^\infty\big(-1+{\pi^2n^2\over L^2}\big)\right|}}\nonumber \\
&=&
\frac{1}{2^{3/4}\pi}\sqrt{\frac{\sinh(\sqrt2L)}{\sin L}}\,,
\end{eqnarray}
where the latter expression follows from the well-known infinite product
representation for the sine.  This diverges as $L\to\pi^-$, i.e., as $L\to
(L^N_c)^-$.  In this limit, $\Gamma_0\sim{\rm
const}\times(L^N_c-L)^{-1/2}$.  As mentioned in the introduction, this
divergence has a simple physical interpretation: it arises from the MPEP
becoming {\em transversally unstable\/} as $L\to (L^N_c)^-$.  In the
endpoint linearization framework, the divergence is caused by the
eigenvalue $\lambda_{1}^u$ tending to zero.  The appearance of a zero
eigenvalue signals the appearance of a transverse soft mode.

\subsection{The Neumann-Case Rate Prefactor When $L>L_c$}
\label{subsec:supercrit}

When $L>L_c^N$, there are two transition states: the nonuniform droplet
pair configurations $\pm\phi_u$, one of which is shown in
Fig.~\ref{fig:sols}.  Closed-form computation of the eigenvalues of the
linearized dynamics at a nonuniform stationary state is generally not
possible.  However, there are well-known techniques for computing the
determinant quotient in~(\ref{eq:detrat}).  The transition state~$\phi_u$
is given by~(\ref{eq:neuinst}), with $m=m(L)$ (for ease of notation, all
subscripts will be dropped) determined implicitly by~(\ref{eq:neuu}).  The
associated linearized evolution operator, computed from~(\ref{eq:phi0}), is
\begin{equation}
\label{eq:phiuop}
\hat\Lambda[\phi_u]= -{d^2\over dx^2}-1+{6m\over m+1}\ {\rm
sn}^2\left({x\over\sqrt{m+1}}+{\bf K}(m)\ \Big|\ m\right)\, .
\end{equation}
Evaluation of $\Gamma_0$ requires the calculation of the eigenvalue
spectrum of this unusual Schr\"odinger operator, or at~least the
calculation of an associated determinant quotient.  It~is worth mentioning
that a similar Schr\"odinger operator arises in the analyses of Langer and
Callan--Coleman.  However, since they considered only the limit
$L\to\infty$, which corresponds to $m\to1^-$, the `potential energy' in
their Schr\"odinger operators involves hyperbolic trigonometric functions
rather than elliptic functions.  For a careful review, see
Ref.~\cite{Schulman}.

Calculation of the associated determinant quotient is facilitated by the
following fact, which was first noticed by Gel'fand (c.~1960).  Let
$\eta_{u,*}$ be a nonzero solution on $[0,L]$ of
the homogeneous equation
\begin{equation}
\label{eq:homog}
\hat\Lambda[\phi_u]\eta=0\,,
\end{equation}
and let $\eta_{s,*}$ be chosen similarly.  Then if $\eta_{u,*},\eta_{s,*}$
satisfy a Neumann boundary condition at~$x=0$, i.e.,
\begin{equation}
\label{eq:bcs}
\eta'_{u,*}|_0=\eta'_{s,*}|_0=0\,,
\end{equation}
it can be shown that~\cite{Forman87,McKane95}
\begin{equation}
\label{eq:ratio}
{\det\hat\Lambda[\phi_s]\over\det\hat\Lambda[\phi_u]}={\eta'_{s,*}|_L\,\eta_{u,*}|_0\over\eta'_{u,*}|_L\,\eta_{s,*}|_0}\,.
\end{equation}
This reduces the calculation to a series of manipulations of solutions of
homogeneous Schr\"odinger equations.

Consider first the ratio $\eta'_{u,*}|_L\Big/\eta_{u,*}|_0$.  A~solution of
the homogeneous equation~(\ref{eq:homog}) that satisfies the Neumann
boundary conditions can be found (see Ref.~\cite{McKane95}) by
differentiating the periodic instanton solution~(\ref{eq:neuinst}) with
respect to~$m$; the result is
\begin{eqnarray}
\label{eq:eta2} 
\eta_{u,*}(z;m)&=&{1+m^2\over(1-m)\sqrt{2m(1+m)^3}}\,{\rm
sn}(z|m)-\sqrt{m\over2(1+m)}{1\over 1-m}\,{\rm sn}^3(z|m)\\
&&\quad{}+{1\over\sqrt{2m(1+m)^3}}\,{\rm cn}(z|m)\,{\rm dn}(z|m)\left[z-{\bf K}(m)+{1+m\over 1-m}\Big({\bf E}(m)-{\bf E}(z|m)\Big)\right]
\,,
\nonumber
\end{eqnarray}
where $z\equiv x/\sqrt{m+1}+{\bf K}(m)$, and ${\rm cn}(\cdot\mid m)$ and
${\rm dn}(\cdot\mid m)$ are Jacobi elliptic functions~\cite{Abramowitz65}.
This yields
\begin{equation}
\label{eq:unstable}
\eta'_{u,*}|_L\Big/\eta_{u,*}|_0=2\left[(1-m){\bf K}(m)-(1+m){\bf E}(m)\right]\, .
\end{equation}
The other ratio in~(\ref{eq:ratio}) is easily computed by choosing
$\eta_{s,*}=\cosh(\sqrt{2}x)$, which yields
\begin{equation}
\label{eq:stable}
\eta'_{s,*}|_L\Big/\eta_{s,*}|_0=\sqrt{2}\sinh(\sqrt{2}L)\, .
\end{equation}
This is consistent with the numerator of~(\ref{eq:g0-}), which was obtained
through direct computation of the eigenvalue spectrum.  Substituting
(\ref{eq:unstable}) and~(\ref{eq:stable}) into~(\ref{eq:ratio}) yields a
formula for the determinant quotient.

Finally, we compute the single negative eigenvalue $\lambda_{0}^u$
of~$\hat\Lambda[\phi_u]$, which is associated with downhill motion away
from the transition state~$\phi_u$.  It is easy to check that the
corresponding eigenfunction is
\begin{equation}
\label{eq:eigenvector}
\eta_{u,0}(x;m)={\rm sn}^2\left({x\over\sqrt{m+1}}+{\bf K}(m)\ \Big|\ m\right)-{1\over 1+m-\sqrt{m^2-m+1}}\, .
\end{equation}
In physical terms, this eigenfunction specifies the way in which a droplet
pair configuration in the Neumann model begins to collapse, due to the
boundary between the positive and negative droplets, which is a kink
initially localized at~$x=L/2$ (see Fig.~\ref{fig:sols}) tending to move
toward $x=0$ or~$x=L$, where it will be annihilated.  The negative
eigenvalue itself is
\begin{equation}
\label{eq:eigenvalue}
\lambda_{0}^u=1-{2\over 1+m}\sqrt{m^2-m+1}\, ,
\end{equation}
which approaches $-1$ as $m\to 0^+$, i.e., as $L\to(L_c^N)^+$, in agreement
with the single negative eigenvalue of~(\ref{eq:unstablespectrum}).  As
$m\to 1^-$ (i.e., as $L\to\infty$), $\lambda_{0}^u\to 0$ as ${\rm
const}\times \exp(-L\sqrt{2})$.  This implies that the zero-noise motion of
the kink domain wall between the positive and negative droplets, which,
since it leads to annihilation of the kink, is a deterministic `coarsening'
phenomenon (cf.~Ref.~\cite{Habib2000}), proceeds only very slowly in the
limit of large~$L$.  In the same limit, the numerator of the determinant
quotient diverges only as $\exp(L/\sqrt{2})$, so the Kramers rate prefactor
tends to zero.

As an interesting aside, we note that the eigenvalue equation for the
operator $\hat\Lambda[\phi_u]$ of~(\ref{eq:phiuop}) is the spin-$2$ Lam\'e
equation~\cite{Li00}, which is a Schr\"odinger equation with an elliptic
potential.  This potential is periodic, with lattice constant~$2{\bf
K}(m)$.  So the spectrum of~$\hat\Lambda[\phi_u]$ has a band structure.
This observation facilitates the calculation of the eigenvalues
of~$\hat\Lambda[\phi_u]$, which can be viewed as band edges; and, though we
do not supply details here, the calculation of associated determinant
quotients.  The fact that the Lam\'e equation is the eigenvalue equation
that governs the stability of periodic instanton configurations has
previously been noticed by others~\cite{Liang92,Caputo2000}, though mostly
in a quantum context.

Putting everything together, we find the Neumann-case rate prefactor when
$L>L_c^N=\pi$ to be
\begin{equation}
\label{eq:prefabove}
\Gamma_0= {1\over \pi} \left|1-{2\over 1+m}\sqrt{m^2-m+1}\right|
\sqrt{{{\sinh(\sqrt{2}L)}\over{\sqrt{2}\left|(1-m){\bf K}(m)-(1+m){\bf E}(m)\right|}}}
\, .
\end{equation}
Compared to the formula~(\ref{eq:detrat}) for~$\Gamma_0$, this includes an
extra factor of~$2$, since there are two transition states: $\phi_u$
and~$-\phi_u$.  As~$m\to 0$ (i.e., as $L\to (L_c^N)^+$),
$\Gamma_0$~diverges as ${\rm const}\times(L-L_c^N)^{-1/2}$.  It~should be
noted that as $L\to (L_c^N)^+$ and $L\to (L_c^N)^-$, the prefactor diverges
according to the same inverse power law.

The divergence at $L=\pi$ is certainly striking, and leads one to wonder
how, when $L=\pi$ exactly, the Kramers rate formula
$\Gamma\sim\Gamma_0\exp(-\Delta W/\epsilon)$ should be modified.
Presumably, it should be replaced by
\begin{equation}
\Gamma\sim{\rm const}\times \epsilon^{-\alpha} \exp(-\Delta W/\epsilon)\,,\qquad\epsilon\to0\,,
\end{equation}
for some $\alpha>0$.  The computation of the exponent~$\alpha$ and the
factor `${\rm const}$', and of~course the divergence of the rate prefactor
at $L=L_c$ for the other choices of boundary condition, will be considered
elsewhere.

\section*{Acknowledgments}
 
This research was supported in part by National Science Foundation Grant
No.~PHY-0099484.


\end{document}